\documentstyle[sprocl]{article} 

\def\|{\'{\i}}

\def\be{\begin{equation}}\def\ee{\end{equation}}\def\bea{\begin{eqnarray}}
\def\eea{\end{eqnarray}}

\newcommand{\eps}{\varepsilon}

\begin{document}

\title{\bf Relativistic Bound States in 2+1 and 1+1 
Dimensions in the Null-Plane 
\footnote{To appear in "Proceedings VI Hadrons 1998", Florian\'opolis,
Santa Catarina, Brazil}}
\author{ J. P. B. C. de Melo$^{(a)}$,
A. E. A. Amorim$^{(b)}$, Lauro Tomio$^{(c)}$ and T.Frederico$^{(d)}$
\\ $^{(a)}$ Division de Physique Th\'eorique, 
Institut de Physique Nucl\'eaire, 
91406 Orsay CEDEX,
and Laboratoire de Physique Th\'eorique et Particules El\'ementaires, \\ 
Universt\'e Pierre et Marie Curie, 4 Place Jussieau, 75252 
Paris CEDEX 05, France
\\ $^{(b)}$ 
Faculdade de Tecnologia de Jahu, CEETPS, Jahu, Brasil.
\\ $^{(c)}$
Instituto de F\'\i sica Te\'orica, UNESP \\
01405 S\~ao Paulo, SP, Brasil \\
$^{(d)}$
Dept. de F\'\i sica,
Instituto Tecnol\'ogico da Aeron\'autica,CTA \\
12.228-900 S. Jos\'e dos Campos, SP, Brasil}
\vspace{2 true cm}
\date{}
\maketitle
\vspace{0.5 true cm}

The Faddeev-like equation for the component of the three-boson
vertex for a relativistic contact interaction is \cite{fred92} 
\begin{eqnarray}
v(q^\mu)=2\tau(M_2) \int \frac{d^nk}{(2\pi)^n}
\frac{i}{k^2-M^2+i\eps} \frac{i}{(P_3-q-k)^2-M^2+i\eps}v(k^\mu),
\label{1}
\end{eqnarray}
where $P^\mu_3=(M_{3B},0,0,0)$ is the three-boson four-momentum in the
center of mass system, $n$ is the dimension of the space-time,
the mass of the two-boson subsystem is given by
$M_2^2=(P_3-q)^2$ and the single boson mass is $M$. 
The factor 2 comes from symmetrization of the total vertex.
 
The total three-boson vertex is the sum of  three Faddeev components 
in which each boson is spectator once \cite{fred92}.
The two-boson scattering amplitude, $\tau(M_2)$, 
enters in the kernel of the 
integral equation for the vertex in Eq.(\ref{1}). It is 
easily obtained as: 
\begin{eqnarray} 
\tau^{(n)}(M_2)= \left\{i\lambda^{-1} - B^{(n)}(M_2)\right\}^{-1}, 
\label{2}
\end{eqnarray}
where $\lambda$ is the coupling constant of the zero-range interaction,
$M_2$ is the  mass of the two boson system and
$B(M_2)$ is the kernel of the integral equation
for the scattering amplitude
\begin{eqnarray} B^{(n)}(M_2)=-\int \frac{d^nk}{(2\pi)^n}
\left\{\left(k^2-M^2+i\eps\right)\left((P-k)^2-M^2+i\eps\right)
\right\}^{-1},
\label{3}
\end{eqnarray}
where $M$ is the boson mass and $P^\mu=(M_2,0,0,0)$ .

The value of 
$\lambda$ is chosen such that the two-boson system has
one bound-state. The scattering
amplitude, Eq.(\ref{2}), has a pole
at the bound-state mass $(M_{2B})$, which yields
\begin{eqnarray} 
i\lambda^{-1}= B^{(n)}(M_{2B}). 
\label{4}
\end{eqnarray}

In two dimensions the two-boson
scattering amplitude for $M_2 \ < \ 2M$ is
\begin{eqnarray}
\tau^{(2)}(M_2) = - 2\pi i \left\{ 
\frac{atan\left(2\beta(M_{2B})\right)^{-1}}
{M^{2}_{2B}\beta(M_{2B})}
 - \frac{atan\left(2\beta(M_{2})\right)^{-1}} 
{M^{2}_{2}\beta(M_{2})}
\right\}^{-1} \ ,
\label{5}
\end{eqnarray}
where
$ \beta(M_2)=\sqrt{\frac{M^2}{M_{2}^2}-\frac14}$ .

In three dimensions the scattering amplitude is
\begin{eqnarray} 
\tau^{(3)}(M_2)=- 8\pi i \left\{ M^{-1}_{2B} 
ln\left(\frac{2M+M_{2B}}{2M-M_{2B}} 
\right)
-M^{-1}_{2} ln\left(\frac{2M+M_{2}}{2M-M_{2}} 
\right)
\right\}^{-1} \ .
\label{6} 
\end{eqnarray}
For our purpose of the bound-state calculation
is enough to know $\tau^{(n)}(M_2)$ for $M_2 < 2M$ .

The momentum
variables in the integral equation
are the momenta in the null-plane for an on-mass-shell particle,
$q^+$ and $q_\perp$ \cite{fred92}. 
The transversal momentum is needed in three space-time dimensions. 

Let us discuss the limits of the variables
$y=\frac{q^+}{M_{3B}}$ and $q_\perp$.
In 1+1 space-time dimensions only the momentum fraction
is enough to describe the spectator boson.
 The mass of the two-boson
subsystem must be real and in 1+1 dimensions it implies
\begin{eqnarray} 
(M_2)^2\ = \ (M_{3B}-q^+)\left(M_{3B}-\frac{
M^2}{q^+}\right) \ > \ 0 \ . 
\label{7}
\end{eqnarray}
From the above inequality follows
$ 1\ > \ y \ > \ \frac{M^2}{M^2_{3B}} \ . $

In 2+1 dimensions, we deduce the range of values 
of the perpendicular momentum allowed by 
the reality of the mass of the two-boson subsystem. 
Then
\begin{eqnarray} 
(M_2)^2\ = \ (M_{3B}-q^+)\left(M_{3B}-\frac{q^2_\perp
+M^2}{q^+}\right)-q^2_\perp \ > \ 0 \ . 
\label{8}
\end{eqnarray}
Solving the inequality for $q^2_\perp$ , we obtain
$ q^2_\perp\ < \ (1-y)(M_{3B}^2y-M^2) \ . $
The limits for $y$ are
$ 1 \ > \  y \ > \ \frac{M^2}{M^2_{3B}} \ , $ and
the lower bound comes from $q^2_\perp \ > \ 0$.

The equation for the Faddeev component of the
vertex in 1+1 space-time dimensions is obtained
as the result of the $k^-$ integration in the momentum loop 
of Eq.(\ref{1}). We also use Eq.(\ref{5}) and the 
limit in the internal momentum fraction $x$ 
\begin{eqnarray}
v(y)=\frac{i}{2\pi}\tau^{(2)}(M_2) \int^{1-y}_\frac{M^2}{M^2_{3B}}\frac{dx}{x(1-y-x)}
\frac{v(x)}{M^2_{3B}-M^2_{03}},
\label{9}
\end{eqnarray}
where $(M_2)^2$ is given by Eq.(\ref{7})
and the free mass of the virtual three-boson state in 1+1 dimensions
is:
\begin{eqnarray} 
M^2_{03}=
\frac{M^2}{x}+
\frac{M^2}{y} + \frac{M^2}{1-y-x} \ .
\label{10}
\end{eqnarray}

The equation for the Faddeev component of the
vertex in 2+1 space-time dimensions 
is found after the $k^-$ integration of in the momentum loop of
Eq.(\ref{1}),
\begin{eqnarray}
v(y,\vec q_\perp)=\pi^{-2}
\tau^{(3)}(M_2)
\int^{1-y}_\frac{M^2}{M^2_{3B}}\frac{dx}{x(1-y-x)}
\int^{k_\perp^{max}}_{-k_\perp^{max}}d^2k_\perp
\frac{v(x,\vec k_\perp)}{M^2_{3B}-M^2_{03}},
\label{11}
\end{eqnarray}
where $M_2$ is given by Eq.(\ref{8}), as well as
$ k_\perp^{max}=\sqrt{(1-x)(M_{3B}^2x-M^2)}$ .
The mass of the virtual three-boson state is:
\begin{eqnarray} M^2_{03}=
\frac{k^2_\perp+M^2}{x}+
\frac{q^2_\perp+M^2}{y} + \frac{(q+k)^2_\perp +M^2}{1-y-x} \ .
\label{12}
\end{eqnarray}

The dependence of $v$ on $q^-$ is not specified because the
spectator boson is on mass-shell. $q^+$ and $\vec q_\perp$
describe the spectator boson propagation.
The relativistic equations in 1+1 and
2+1 dimensions, have a lower
bound for the mass of the three-boson system which comes 
from the limits on the $x$ integration and the condition 
$y>\frac{M^2}{M^2_{3B}}$ which implies  $ M_{3B}> \sqrt{2}M \ . $
The same limit was obtained in 3+1 dimensions in \cite{fred92}.

The Faddeev component of the ground state vertex, in 2+1 dimensions 
is rotationally symmetric in the x-y plane in Eq.(\ref{11}). 
The mass of the
boson gives the scale of the system. Here, the solution is presented for
 $M=1$.
In Fig.(1), the numerical results for the ground-state binding energies 
$(E_{3B}=M_{2B}+M-M_{3B})$ of
the three-boson system are shown in two and three-dimensions. 
In the nonrelativistic limit, for $E_{2B}=0$, the results approach
the well-known values \cite{dodd}. 

In summary, we give an example of how  null-plane dynamics can be elaborated.
We develop a zero-range model of the three-boson bound state
in the null-plane and  solve numerically the dynamical equation for the
ground-state in 1+1 and 2+1 space-time dimensions.

\vspace{1.0 cm}

This work was supported by Conselho Nacional de Desenvolvimento 
e Pesquisa  - CNPq and Funda\c{c}\~ao de Amparo a pesquisa do 
Estado de S\~ao Paulo - 
FAPESP. J. P. B. C. de Melo is a FAPESP-Brazil fellow 
(contract 97/13902-8).

\begin{figure}[ht]
\vspace{12.0cm}
\caption{The ratio between the three-boson  and two-boson binding energies
as a function of the two-boson binding energy 
in units of the mass of the
single boson. Results for 2+1 (solid line)
and 1+1 (dashed line).}  
\includegraphics{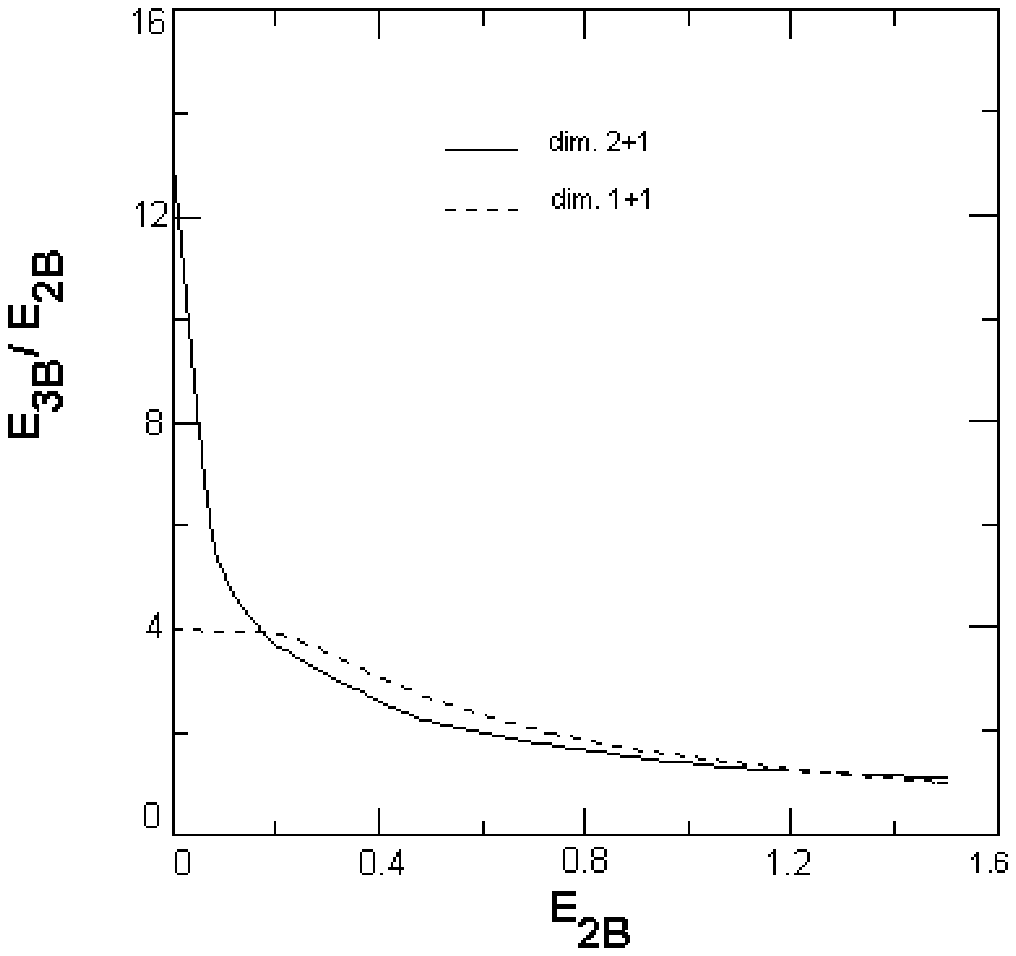}  
\label{fig1} 
\end{figure}

\end{document}